# BEYOND SPACE AND TIME
*The secret network of the universe: How quantum geometry might complete Einstein's dream*

By Rüdiger Vaas

---

With the help of a few innocuous - albeit subtle and potent - equations, Abhay Ashtekar can escape the realm of ordinary space and time. The mathematics developed specifically for this purpose makes it possible to look behind the scenes of the apparent stage of all events - or better yet: to shed light on the very foundation of reality. What sounds like black magic, is actually incredibly hard physics. Were Albert Einstein alive today, it would have given him great pleasure. For the goal is to fulfil the big dream of a unified theory of gravity and the quantum world. With the new branch of science of quantum geometry, also called loop quantum gravity, Ashtekar has come close to fulfilling this dream - and tries thereby, in addition, to answer the ultimate questions of physics: the mysteries of the big bang and black holes.

"On the Planck scale there is a precise, rich, and discrete structure," says Ashtekar, professor of physics and Director of the Center for Gravitational Physics and Geometry at Pennsylvania State University. The Planck scale is the smallest possible length scale with units of the order of $10^{-33}$ centimeters. That is 20 orders of magnitude smaller than what the world's best particle accelerators can detect. At this scale, Einstein's theory of general relativity fails. Its subject is the connection between space, time, matter and energy. But on the Planck scale it gives unreasonable values - absurd infinities and singularities. It carries therefore - as the American physicist John Wheeler, who knew Einstein personally, used to say - the seeds of its own destruction. That means the theory indicates the limitations of its own applicability. This is a restriction, but at the same time also an advantage: physicists cannot avoid looking for a better and more complete theory for the laws of nature at this fundamental level. In other words: they need a theory of quantum gravity in order to explain the behavior of nature at all levels, from quarks to quasars.

Yet nature so far seems to follow two different types of rules:
 - On the one hand general relativity: its alphabet is geometry, and its vocabulary consists of lines, angles, surfaces and curves. Gravity is a property of the geometry of a spacetime that is
not merely a stage of all events, but also a participating actor.
 - On the other hand quantum theory: its alphabet consists of algebraic symbols and quantum numbers and does not contain the deterministic words "always" and "never", but rather the statistical ones "usually" and "rarely". Here spacetime is a fixed stage ("background metric") for particles and forces. Gravitation is described by - hypothetical, not yet proven - particles called gravitons. They are exchanged in a subatomic ping-pong between all other particles and create the force of gravity. But since the gravitons also interact with themselves, the tools of quantum physics, which are so successful elsewhere, run into severe difficulties due to absurd infinities, probabilities over 100 percent, and other problems.

Therefore both theories are not compatible on a fundamental level. Albert Einstein had recognized that already. And since that time theoretical physicists have looked for a synthesis - a "theory of everything" as some immodestly call it. It would contain the fundamental laws of nature - the code or blueprint of our universe.

Most attempts in this area are undertaken by particle physicists assuming a flat background spacetime. "If we remove life from Einstein's beautiful theory by steam-rollering it first to flatness and linearity, then we shall learn nothing from attempting to wave the magic wand of quantum theory over the resulting corpse," criticizes University of Oxford's mathematician Roger Penrose. The equations describing the behavior of gravity under quantum conditions cannot be solved, although they seem meaningful and consistent. It is like a palace that has no entrance.

Relativists prefer to approach the problem from a geometrical point of view. John Wheeler speculated already in the 1950s that on smallest scales spacetime is no longer homogeneous but rather "foamy". This is a metaphor that does not claim any scientific precision. It is clear however that a theory of quantum gravity must entail powerful revisions of our worldview and would stretch our intuition from everyday life beyond its limits.

"There is no more challenging problem in science than the completion of this theory. It will provide new answers to the questions of what space and time are. But that is not all. It will also be a theory of matter and a theory of cosmology," says Ashtekar's colleague Lee Smolin. "Presently we are in a crucial period during which the laws of physics are being rewritten." The professor of physics at the Canadian University of Waterloo has no doubt about the radical consequences. "The continuous appearance of space is as much an illusion as the smooth appearance of matter. Were we able to observe the world on a small enough scale, we would see that space is made of things that we can count."

This is how quantum geometry revolutionizes our worldview: Space is quantized like matter!

The question why nothing can be made to fit into half the volume of the smallest unit of space is meaningless in view of such "space atoms". It is based upon the incorrect presumption of an absolute space in which things fit in. Space and time are not at all fundamental, but rather built up from more basic structures. Ashtekar and his colleagues call them spin networks. This concept was coined by Penrose, who already in the 1970s had formulated his Twistor theory with similar motivations, and introduced spin networks as a kind of spacetime dust. Ashtekar compares the spin networks - mathematically known as graphs - to a fabric of polymer-like one-dimensional threads. If one could observe nature with maximum possible enlargement, space and time would dissolve and the granular mesh of the spin network would come to light - or more precisely: the quantum physical superposition of all possible configurations of these entities. There is 'nothing' between these graphs. Those entities rest only on themselves, so to speak. "The spin networks do not exist in the space. Their structures produce the space," Smolin stresses. "And they are nothing but abstractly defined relations - which determine how the edges come together and interlock at the joints ."

That space appears nevertheless homogeneous to us is no miracle. For, the resolution of our perception is limited, similar to observing a photograph whose individual pixels we cannot recognize from a distance. With the "small" difference that there are, believe it or not, $10^{68}$ quantum threads that intersect the page of an ordinary physics journal.

The endpoints of the graphs (i.e., open graphs) represent fermions (quarks and leptons), out of which matter is made, and the presumed Higgs boson that gives them mass. The bosons (such as photons and also gravitons), which transmit the subatomic interactions and thus evoke the forces of nature, are encoded in certain excited states of the spin network as changing colors or labels on the graphs. Ashtekar: "Some represent geometry, others fields. Matter can only live where geometry is excited, and that is only along the polymers. It is physically not meaningful to ask what is between the edges of the graphs. Gravitons and so on are not fundamental, but rather a product of the spin networks." Our usual notion of causality does not apply here either. Even time only emerges from the variations of the excited states and the links in the networks. In a certain way time is an illusion just as space is. Ashtekar quotes Vladimir Nabokov: "Space is a swarming in the eyes, and Time a singing in the ears."

The entire realm of reality therefore originates from the superposition of fluctuating weaves on a submicroscopic level. We, and everything we know, are like patterns or embroideries in the fabric of the spin network. With a twinkle in his eye, Ashtekar even took a few weaving lessons to understand better what he is talking about.

The pleasant scientist with the spark of enthusiasm in his eyes was born in 1949 in the small town of Shirpur in western India. As a youth, he read the popular science books of the Russian-American cosmologist George Gamow and wanted to become a physicist. His aptitude already showed itself early when he found a small mistake in a textbook of the Nobel Prize winner Richard Feynman. He wrote him. "Feynman actually answered and acknowledged I was right. That was so uplifting for me, that I still have the letter," says Ashtekar. He studied physics in Bombay and from 1969 in the USA, and held positions in Oxford, Chicago, Paris and Syracuse, before he came to Penn State University. In Germany and Austria, he also worked as a guest scientist and was even offered a position as Director of the Max Planck Institute for Gravitational Physics in Potsdam. "That was a great honor for me. But Penn State thereupon offered me much more time and freedom for my research than I would have had in Germany, and so I had to regretfully decline."

Already early in his career, Ashtekar had turned towards quantum gravity. "There is a kind of innocent arrogance when one is young, about working on the most difficult problems."

The first pillars of the bridge between general relativity and quantum theory were erected in 1986. Inspired by a paper on the motion of electrons in the gravitational field - written by Amitabha Sen, then a student at the University of Chicago - Ashtekar developed a new geometric language, in which Einstein's field equations could be formulated differently but in a mathematically equivalent manner. This version quickly found extensive worldwide recognition. It resembles the more easily manageable equations of the electro-weak interaction and Maxwell's equations. In this way gravity got a more familiar form. But certainly only for the experts, for fundamental ideas such as 'flux variable', 'connection', and 'holonomy' evoke only a frown in a layman. At the same time it allows elegant methods to describe places, areas, motion and forces without requiring a background metric. Other quantities are already established as "Ashtekar variables" in the relevant textbooks.

But that was not enough. In the next step, after laborious and detailed work, Ashtekar's version of Einstein's equations was successfully extended so that they could be quantized. "What we found exceeded our wildest expectations," remembers Lee Smolin. Together with the Italian physicist Carlo Rovelli, currently at the University of Marseille, he performed the decisive pioneering work between 1988 and 1990. Then in 1992 they collaborated with Ashtekar. On this level of description, space would no longer look homogeneous, but rather have a fine-grained structure. It resembled chain mail, the medieval knight's armor, constructed out of innumerable multiply connected rings ("loops"), each only one Planck-length small. The theory of loop quantum gravity was born.

If an atom were as large as the Milky Way, quantum loops would be no larger than a human cell. "It is therefore not surprising that space looks so smooth, similar to a knitted T-shirt which looks evenly smooth from a large distance," says Rovelli. For fun he had fashioned a model from metallic key rings to give the abstract vision a vivid clarity - "out of each ring to be found in Verona," he says. "I bought out all the shops." At that time he was a type of mediator between the quiet, contemplative, analytical physicist, Ashtekar, who loved Mozart and demanding literature and philosophy, and had a stunningly orderly office, and the restless almost chaotic-creative Smolin whose office looked as if a hurricane had mixed up all the books, periodicals, and clothes therein.

A crucial source of inspiration was the so-called Wilson-loops in the lattice gauge theory of quantum chromodynamics. They had been developed independently by the American physicist Kenneth Wilson and the Russian physicist Alexander Polyakov. Quantum chromodynamics describes the behavior of quarks - the building blocks of protons and neutrons - not on a continuous space but on a lattice structure. "A physicist who works without a lattice, is like a trapeze artist without a net," Smolin describes the advantages of this abstract extension. "There is always the danger that an incorrect step has embarrassing consequences." In physics, these catastrophes emerge through confrontations with infinity and absurd mathematical expressions. This happens in all quantum theories that are based on continuous spacetime.

After months of enthusiasm the disillusionment followed. "The mathematics was unclear, one could easily make a fool of oneself," says Ashtekar. For again the troublesome infinities emerged in the calculations. "The loops cannot be regarded any longer as the fundamental representation of reality. They are still a useful description, as much as Wheeler's quantum foam is, but we were missing the correct mathematical setting. Paradigm changes in theoretical physics frequently require a new mathematical arena: Newton's mechanics and theory of gravity required differential calculus, Maxwell's electrodynamics required partial differential equations and analysis; Einstein's general theory of relativity required differential geometry and quantum mechanics required Hilbert space and operator algebra."

Ashtekar did not give up. In five additional hard years he compiled together with Jerzy Lewandowski, John Baez, Chris Isham, Thomas Thiemann and others the tools for quantum geometry where the mathematical field of knot theory was particularly important. Now, in the center are spin networks and graphs, which are like links and intersections of loops, whereas the spins indicate the type and number of the connections. "The precise mathematics exists now," reports Ashtekar. "The infinities have disappeared, we no longer lead ourselves by the nose." The formalism is so efficient that it can be applied not only to general relativity, but also to other theories of gravity such as supergravity. "Above all, now quantum theory and general relativity are truly talking to each other," says Rovelli, pleased with this successful synthesis.

The next big goal is already being aimed at: the connection between the familiar low-energy physics and the more fundamental physics of spin networks. "Shadow states provide a technical bridge here," says Ashtekar. He thinks about a kind of projection of the physical states onto the graphs. It would be a big success if one could derive the well-known physics in detail from quantum geometry. But that is still not all. The extremely busy Ashtekar is also working on a new formulation of quantum theory to generalize it further in order to make it compatible with general relativity and perhaps even solve its troublesome interpretational problems.

The acid test of quantum geometry however takes place at the other extreme: in the description of the big bang and black holes. "Quantum geometry is matured enough now that it can directly address these problems," Ashtekar is pleased to note. "Quantum physics does not stop at the big bang," he is convinced. "The classical spactime dissolves near the big bang, but the spin network is still there." It is to a certain extent eternal. "There was thus no emergence of the universe from 'nothing', because 'nothing' simply does not exist. There was always something already." In this manner quantum geometry has the philosophical advantage of simply getting rid of apparently unsolvable questions. Here its strength of being independent from a background spacetime metric becomes particularly noticeable. "Matter and geometry should both be born together quantum mechanically."

Relevant contributions come from research of Ashtekar's former postdoc Martin Bojowald, now at the Max Planck Institute for Gravitational Physics in Potsdam. He showed how the spin network could have ignited the big bang.

Black holes are at present another central testing area for quantum geometry. Ashtekar has successfully contributed to their understanding within the context of general relativity as well. Recently he discovered how black holes grow. But quantum geometry can explain more - namely, how they shrink again. For black holes are not perfectly black, but rather progressively radiate away over very long periods of time based on quantum mechanical effects. That was Stephen Hawking's spectacular discovery in the year 1974.

"Still we have not yet performed a detailed calculation, deriving the Hawking effect from first principles," says Ashtekar. "However it is possible to do so, although it requires some preliminary work, and we are busy with it." Albert Einstein also showed the way here. "At the beginning of the 20th century he discovered that radiation and matter are not distinct, but rather can transform into each other. In the long run radiation and matter quanta are the same," summarizes Ashtekar. "However we learned from Einstein also that the geometry is a physical entity like matter. Therefore radiation or matter can transform into geometry and vice versa."

That this is indeed the case can now be shown by quantum geometry. Its central statement says: yes, there are quanta of geometry. This is exactly the piece of the puzzle that Hawking was missing. Because his calculations presuppose the classical spacetime of general relativity. Ashtekar: "Hawking did not complete Einstein's vision. He only treated matter and energy quantum mechanically." In quantum geometry however spacetime and so also the event horizon of a black hole are quantized. One can think of its surface as being made out of elementary cells of zeros and ones. Each of these tiny spots corresponds to a "thread" of the spin network that cuts through the horizon surface. There are an unimaginable $10^{77}$ threads in the case of a black hole of one solar mass (and therefore $10^{10^{77}}$ different quantum states, that constitute the enormous entropy of a black hole). The special local characteristics of the network define the horizon. If a black hole evaporates, it loses these threads - similar to losing hair on the head, which becomes even more bald, except that the head does not shrink in the process in contrast to the black hole event horizon. In Hawking radiation, quanta of surface areas are therefore transformed into matter and energy quanta.

"That is what Einstein taught us: geometry is physical. It is even so similar to matter that it can change itself into matter," says Ashtekar, and therefore calls this process Einsteinian Alchemy. The dissolution process does not happen gradually, but rather in discrete steps, therefore quantized – similar to hair falling off only one at a time. "Therefore a black hole does not shrink continuously, but rather behaves like an excited atom, delivering its energy in the famous quantum leaps."

Quantum geometry has yet another consequence that Ashtekar and his colleagues are just beginning to explore: the avoidance of the unphysical singularities at the center of black holes, similar to the case of big bang. Possibly even the notorious information paradox can be solved. "In our part of the universe, information which falls into a black hole is lost, but it reappears in a daughter universe," speculates Martin Bojowald. With a smile he adds: "However only after it passes through a rather uncomfortable quantum phase at the central bounce."

Black holes and the big bang are extremely exotic states. But perhaps there is a chance to test quantum geometry through observations in a less extreme context. Giovanni Amelino-Camelia of the University of La Sapienza in Rome has proposed to study highly energetic photons which travel cosmic distances, for instance from gamma ray bursts or x-ray galaxies. They might show small deviations from classical paths if the light wave is scattered at the discrete knots of quantum geometry. Like the spectrum of an atom, the spectrum of spacetime will not be continuous but quantized.

Until such data exist, quantum gravity is exclusively an arena for theorists. In the meantime about two-dozen research groups worldwide are working on the foundations of quantum geometry. Approximately 2000 scientific articles have already appeared. The recognition among his colleagues is so large that Ashtekar was invited to give a keynote presentation at the TH-2002 conference in Paris. TH stands for "theoretical physics," and the high-caliber international conference - only the fourth since 1953 - had no less a goal than giving, among other things, an overview of the most important branches of contemporary research.

These successes are impressive - and yet the resonance is still small in comparison to string theory, which not only enjoys a larger popularity among theoretical physicists but also has become quite a fashion in the meantime. How this theory delivers, however, is not yet clear.

"The main competitor to quantum geometry is string theory. In contrast to this, Ashtekar's formulation contains no unification of all forces - 'only' gravitation is quantized separately," says Claus Kiefer, professor for physics at the University of Cologne and one of the leading experts in the field of quantum gravity in Germany. "Independent of the correctness of the approach, its value lies in pointing out which aspects in a final theory of quantum gravity are to be expected."

String theory interprets elementary particles as minute oscillating threads and can, in contrast to quantum geometry, describe all four forces of nature. Its disadvantage is that it can be formulated only in nine or ten space dimensions and presupposes a background metric of the classical spacetime of general relativity. To that extent space and time are not even quantized - but this is just what is expected of a complete theory of quantum gravity. Here quantum geometry is therefore ahead.

"Of all the formulations of quantum gravity that I know of, that of Ashtekar is the most promising," Roger Penrose therefore also thinks. "So far string theory does not really agree with the world we see. It requires many complicated assumptions such as extra dimensions and supersymmetry, for which there are no empirical clues, and also provides no definite, univocal prediction for future experiments. All the main problems are unsolved," criticizes Rovelli. "I think it is time to try something else also," he says, advertising for quantum geometry. "Certainly this also has gaps and weaknesses. Like the transition of spin networks to classical spacetime is not yet completely understood, and the calculation of entropy of black holes creates some problems," says Kiefer.

"To be continued," as one might say at the frontier of research. More studies are forthcoming. In any case, there is one thing quantum geometry has shown clearly. As the French author Marcel Proust expressed it: "The best discoveries are not made in new landscapes, but rather when one looks at the world with new eyes."

===============================

**Acknowledgements**
I am very grateful to Abhay Ashtekar, Martin Bojowald, Carlo Rovelli and Amitabha Sen for their great support.

**Further reading**
Lee Smolin: Three Roads to Quantum Gravity. Phoenix , London 2001.
*Brief introduction into quantum gravity:*
http://www.damtp.cam.ac.uk/user/gr/public/qg_home.html
*Homepage of Abhay Ashtekar:*
http://cgpg.gravity.psu.edu/people/Ashtekar/
*Homepage of Carlo Rovelli:*
http://www.cpt.univ-mrs.fr/~rovelli/

```
Translated by Amitabha Sen with permission from
Rüdiger Vaas: Jenseits von Raum und Zeit.
bild der wissenschaft (2003), no. 12 , pp. 50-56.
```